\documentclass[final,5p,times,twocolumn]{elsarticle}

\usepackage{lineno,hyperref}
\modulolinenumbers[5]

\journal{Nuclear Instruments and Methods in Physics Research Section A}

\usepackage{numcompress}

\begin{document}

\begin{frontmatter}

\title{Very high energy gamma-ray astronomy with HAWC}

\author[MPIK]{R. L\'opez-Coto}
\ead{rlopez@mpi-hd.mpg.de}
\author[]{for the HAWC collaboration}

\address[MPIK]{Max-Planck-Institut f\"ur Kernphysik, P.O. Box 103980, D 69029 Heidelberg, Germany}

\begin{abstract}

The High Altitude Water Cherenkov (HAWC) observatory is an air-shower array located in Mexico. It is sensitive to the highest energy photons we detect at the Earth, reaching energies of several tens of TeV. The observatory was completed more than one year ago and we are presenting in this contribution the first results about its performance. We also show the results of the first-year survey, the first flaring events detected by the observatory, its sensitivity to extended sources and the plans for the upgrade that is currently taking place.

\end{abstract}


\end{frontmatter}

\linenumbers

\section{HAWC}

\begin{figure}
\begin{center}
\includegraphics[width=0.49\textwidth]{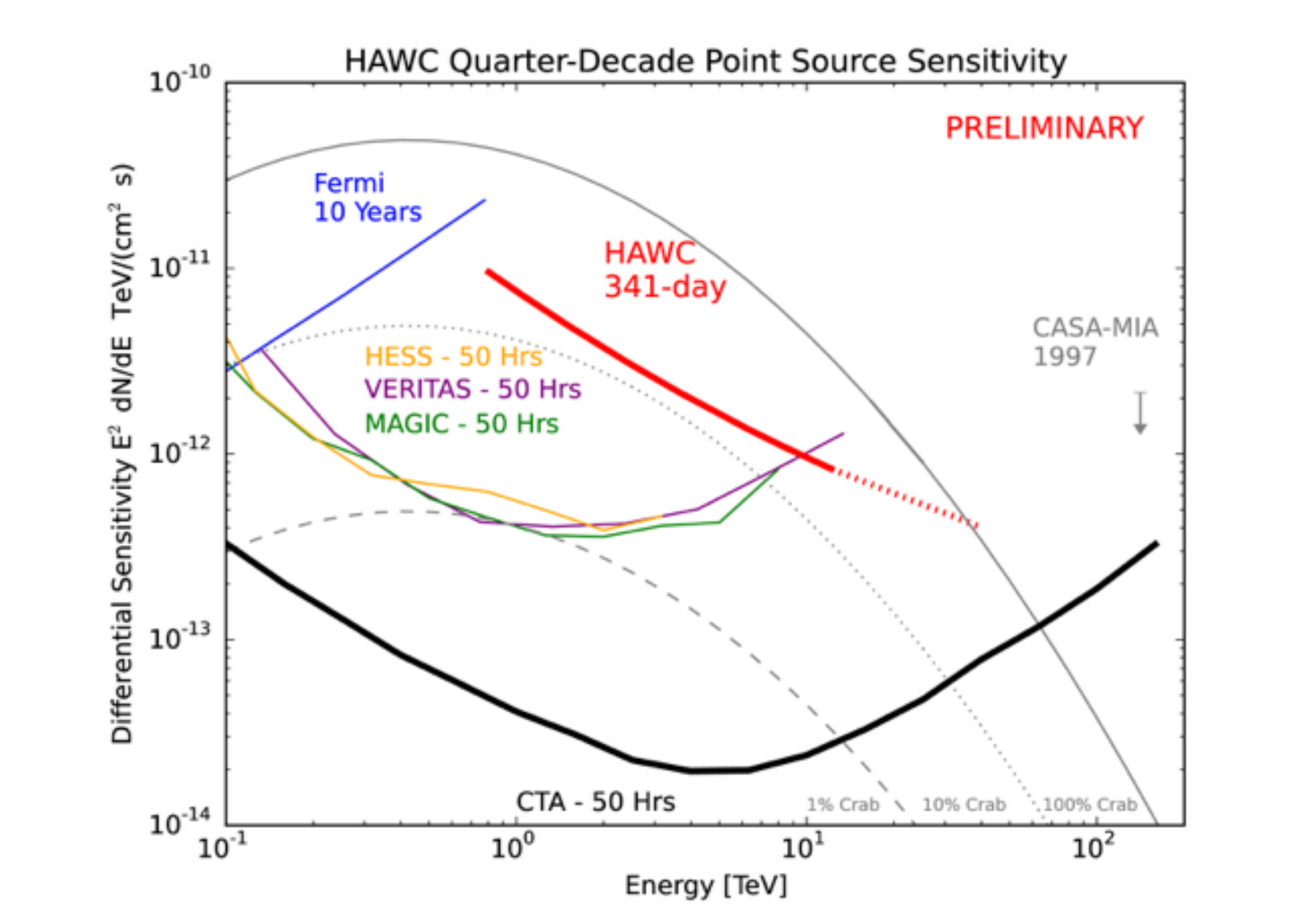}
\caption{Preliminary HAWC sensitivity per quarter decade of energy.}
\label{fig:sens}
\end{center}
\end{figure}

Very High Energy (VHE; E $>$100 GeV) gamma-ray astronomy is a young research field that is producing very interesting results in an electromagnetic window previously unexplored. These highly energetic photons produce a cascade of particles when they interact with the molecules in the atmosphere. The techniques used to detect them are diverse, being the imaging atmospheric Cherenkov technique the one producing the most successful results at multi-GeV energies. For multi-TeV energies, however, the low duty cycle and collection area of these arrays limit the amount of photons that can be detected. The most sensitive technique to detect multi-TeV gamma rays is the Water Cherenkov technique. It makes use of water tanks equipped with photodetectors that measure the Cherenkov light produced by the subparticles from the cascade when they cross the water. The HAWC Gamma-Ray Observatory is located at Sierra Negra, Mexico at 4100 m a.s.l., and is sensitive to gamma rays and cosmic rays in the energy range from 100 GeV to 100 TeV \cite{HAWC_Performance_2013}.
It is composed by 300 optically isolated tanks covering an area of 22 000 m$^2$. Each one of these Water Cherenkov  Detectors  (WCD)  consists  of  a  metallic  cylinder  of  7.3 m  diameter  and  4.5 m  height containing 180 000 liters of water. They are equipped with one 10" PMT at the center and three 8" PMTs surrounding the central one. The array has a 2 sr field of view with $>$95\% uptime. It started operation in its full configuration in March 2015. The current sensitivity curve of HAWC can be seen on Figure \ref{fig:sens}. The scientific results of the partial array are already showing that the instrument is opening a new window to the multi-TeV gamma-ray sky with unprecedented sensitivity. HAWC has just started operation with the full-array completely deployed and the first scientific results are currently being released.

\section{First year survey results}

\begin{figure*}
\begin{center}
\includegraphics[width=0.99\textwidth]{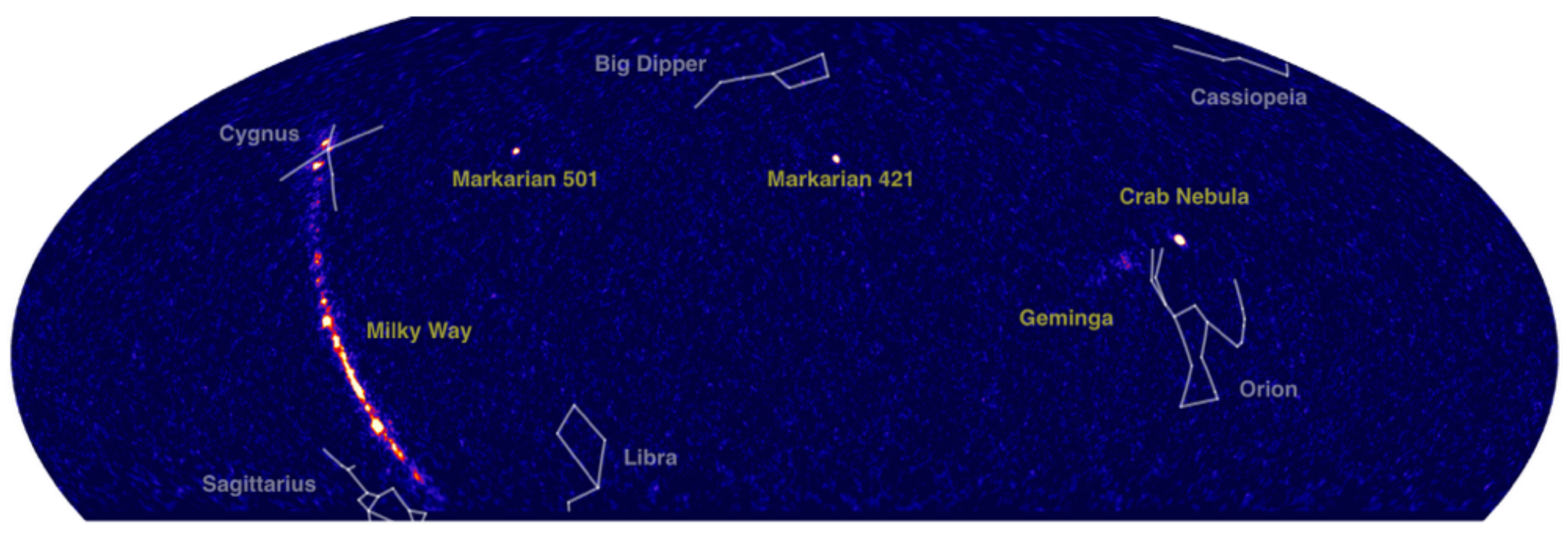}
\caption{First year survey results of the HAWC array.}
\label{fig:survey}
\end{center}
\end{figure*}

After one year of observations, HAWC has detected more than 30 VHE gamma-ray sources, most of them contained in the galactic plane. Several of these sources had already been detected at TeV gamma-rays by imaging atmospheric Cherenkov telescopes, but some of them do not have a counterpart detected by these instruments. A skymap of the 1st year catalog results is shown in Figure \ref{fig:survey}. More details about the survey and the sources detected can be found in \cite{Gamma_Colas}
or in a forthcoming paper.

\section{Sensitivity to flares}

An observatory observing all the sky, taking into account the fast variability of the gamma-ray universe, also targets the monitoring of the sky in order to produce unbiased light curves and to send real-time alerts alerts for pointing instruments with a better sensitivity to observe these phenomena. The HAWC observatory has already detected two AGN flares \cite{Atel1,Atel2}, apart from several non-detected GRBs \cite{Gamma_GRB}.

\section{Extended sources}

\begin{figure*}
\begin{center}
\includegraphics[width=0.9\textwidth]{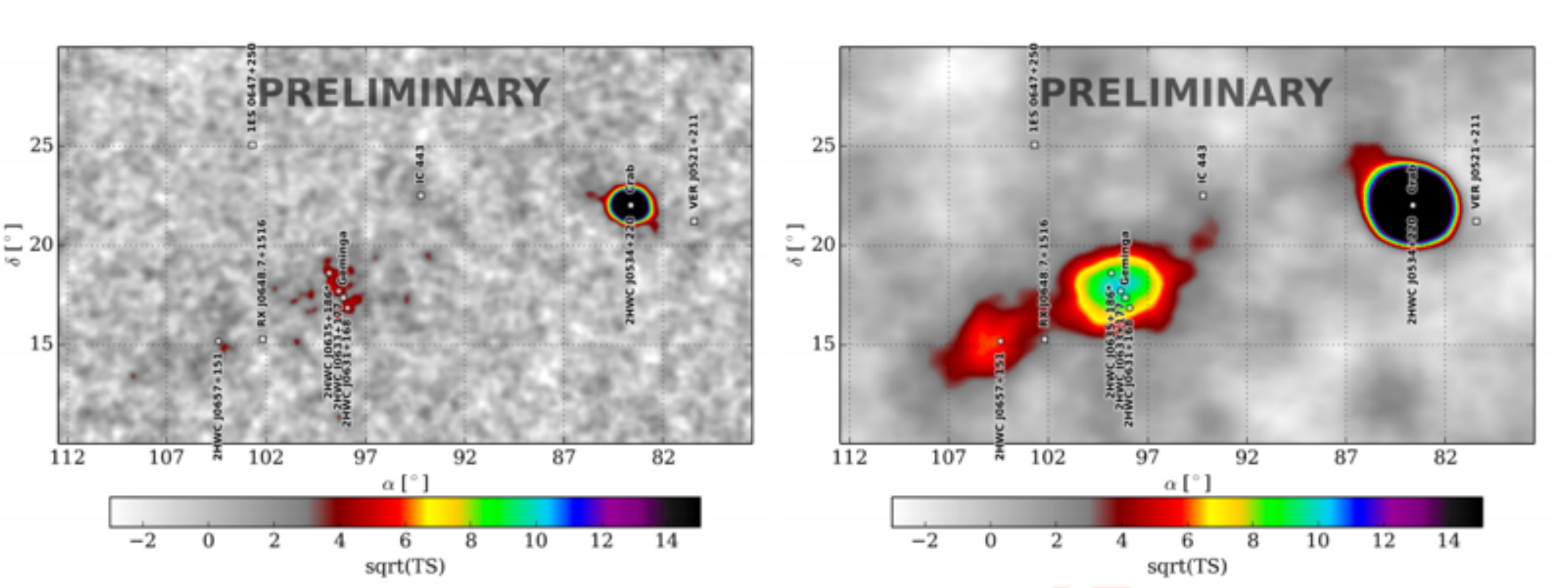}
\caption{Region around Geminga and PSR J0659+14. TS map for point-like source hypothesis (left panel) and TS map for an assumption of a disk of 2.0 degrees.}
\label{fig:extended_sources}
\end{center}
\end{figure*}

One of the advantages of air-shower arrays is their instantaneous 2 sr coverage, that provides a flat background estimation, very important for the detection of very extended sources. HAWC has measured several sources with a pre-trial sensitivity $>5\sigma$ in these searches, but the most prominent are the ones that can be observed in Figure \ref{fig:extended_sources}. There are two sources with extension $>2$ deg surrounding the Geminga and PSR J0659+14 pulsars. These two pulsars have been proposed as the sources of the local flux of cosmic ray e$^\pm$. Using the gamma-ray surface luminosity profile and the spectrum we are measuring the contribution of these sources to the local flux of cosmic ray e$^\pm$. The preliminary results of these studies can be found in  \cite{Gamma_Ruben}
 or in a forthcoming paper.

\section{HAWC Upgrade}
The energy range where HAWC is most sensitive is very important in the search of sources accelerating galactic cosmic rays up to PeV energies. Although the most sensitive instrument at multi-TeV energies, it has a modest good angular and energy resolution due to the difficulty on determining the core position of the shower. HAWC's predecessor, the Milagro experiment, managed to double its sensitivity and halve its angular and energy resolution by installing an outrigger array of smaller water Cherenkov tanks. HAWC is undergoing a similar upgrade, expecting a similar increase in sensitivity, angular and energy resolution at energies above 10 TeV. 

\begin{figure}
\begin{center}
\includegraphics[width=0.4\textwidth]{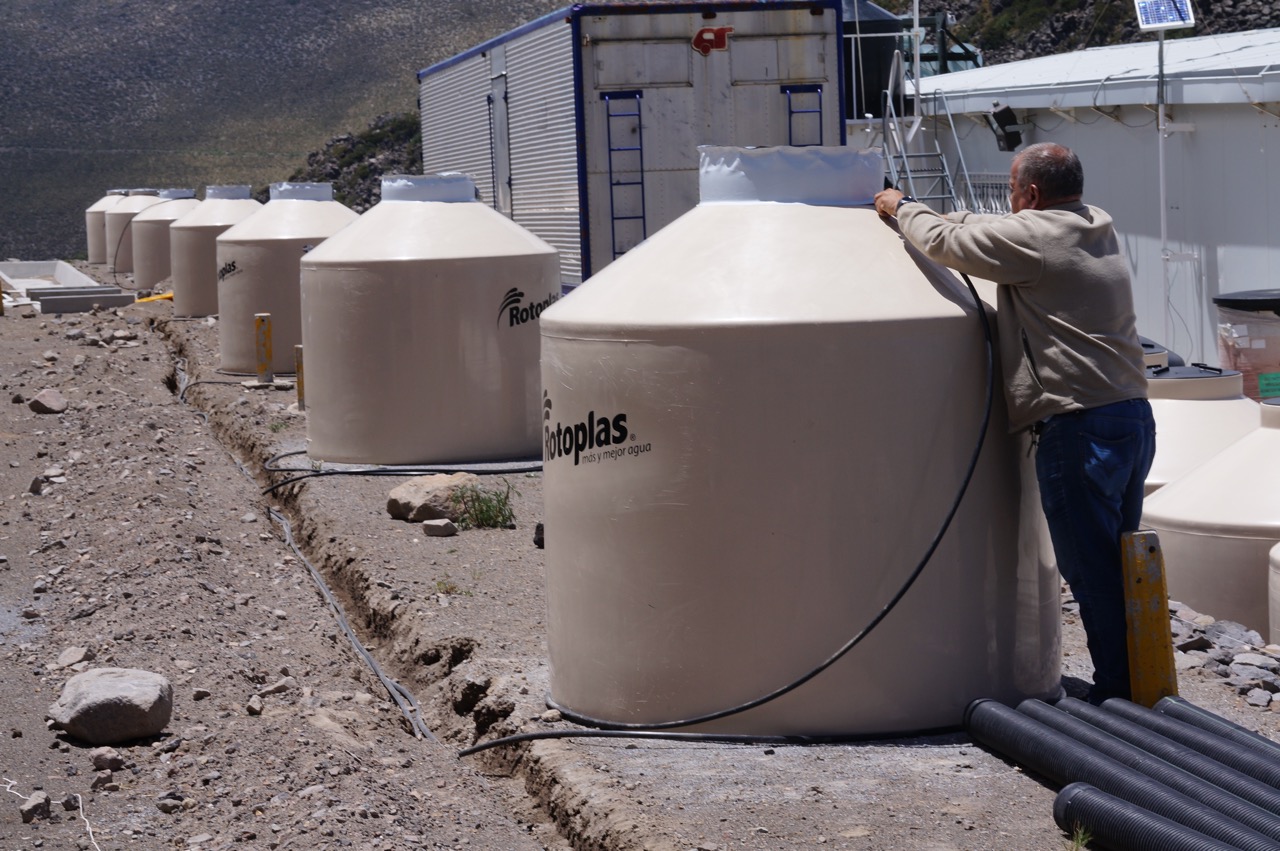}
\includegraphics[width=0.4\textwidth]{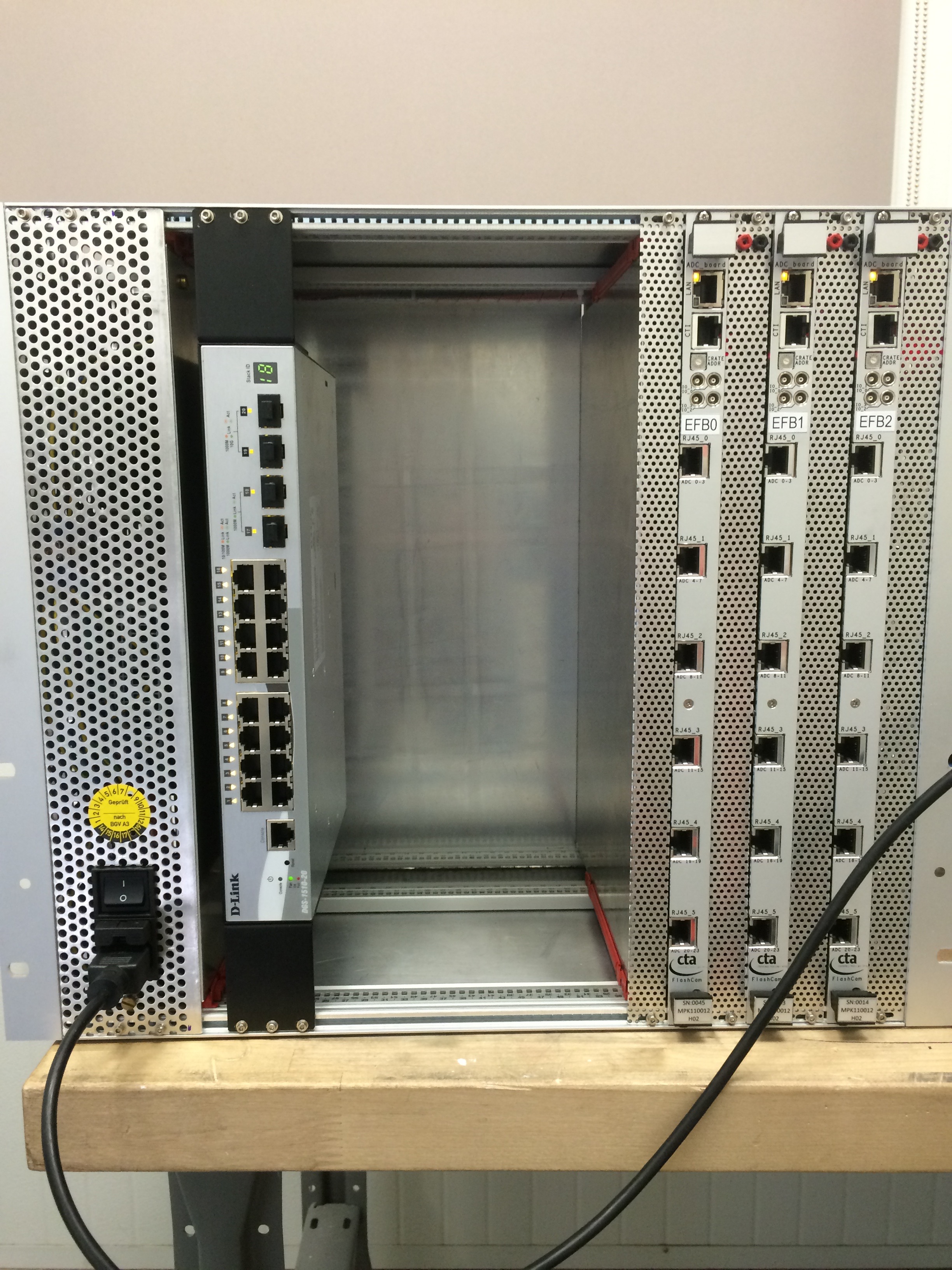}
\caption{Outrigger tanks (top panel) and FADC readout boards (bottom panel).}
\label{fig:outriggers}
\end{center}
\end{figure}

The outrigger array will be composed of 350 cylindrical tanks with a diameter of 1.5 m and height of 1.65 m, with a total volume of 2.5 m$^3$ \cite{HAWC_Outriggers}. A picture of several prototypes for the HAWC outrigger tanks, already installed at the HAWC site is shown in the top panel of Figure \ref{fig:outriggers}. Each one will be equipped with one Hamamatsu R5912 8” PMT, as the ones used to equip 75\% of the channels of the main array. The readout provided is called the Flash ADC eLectronics for the Cherenkov Outrigger Node (FALCON) and can be seen in the bottom panel of Figure \ref{fig:outriggers}. It is based on the design of the FlashCam, one of the readouts for the Medium Size Telescopes of CTA, with some firmware modifications \cite{Flashcam}. 

\section{ACKNOWLEDGMENTS}
We acknowledge the support from: the US National Science Foundation (NSF);
the US Department of Energy Office of High-Energy Physics;
the Laboratory Directed Research and Development (LDRD) program of Los Alamos National Laboratory; 
Consejo Nacional de Ciencia y Tecnolog\'{\i}a (CONACyT),
Mexico (grants 271051, 232656, 55155, 105666, 122331, 132197, 167281, 167733, 254964);
Red de F\'{\i}sica de Altas Energ\'{\i}as, Mexico;
DGAPA-UNAM (grants RG100414, IN108713,  IN121309, IN115409, IN111315);
VIEP-BUAP (grant 161-EXC-2011);
the University of Wisconsin Alumni Research Foundation;
the Institute of Geophysics, Planetary Physics, and Signatures at Los Alamos National Laboratory;
the Luc Binette Foundation UNAM Postdoctoral Fellowship program.

\section*{References}

\bibliography{references}

\begin{thebibliography}{8}
\providecommand{\natexlab}[1]{#1}
\providecommand{\url}[1]{\texttt{#1}}
\providecommand{\href}[2]{#2}
\providecommand{\path}[1]{#1}
\providecommand{\eprint}[1]{\href{http://arxiv.org/abs/#1}{\path{#1}}}
\providecommand{\DOIprefix}{doi:}
\providecommand{\ArXivprefix}{arXiv:}
\providecommand{\URLprefix}{URL: }
\providecommand{\Pubmedprefix}{pmid:}
\providecommand{\doi}[1]{\href{http://dx.doi.org/#1}{\path{#1}}}
\providecommand{\Pubmed}[1]{\href{pmid:#1}{\path{#1}}}
\providecommand{\BIBand}{and}
\providecommand{\bibinfo}[2]{#2}
\ifx\xfnm\undefined \def\xfnm[#1]{\unskip,\space#1}\fi
%Type = Article
\bibitem[{{Abeysekara} et~al.(2013){Abeysekara}, {Alfaro}, {Alvarez},
  {{\'A}lvarez}, {Arceo}, {Arteaga-Vel{\'a}zquez}
  et~al.}]{HAWC_Performance_2013}
\bibinfo{author}{{Abeysekara}\xfnm[ A.U.]}, \bibinfo{author}{{Alfaro}\xfnm[
  R.]}, \bibinfo{author}{{Alvarez}\xfnm[ C.]},
  \bibinfo{author}{{{\'A}lvarez}\xfnm[ J.D.]}, \bibinfo{author}{{Arceo}\xfnm[
  R.]}, \bibinfo{author}{{Arteaga-Vel{\'a}zquez}\xfnm[ J.C.]}, et~al.
\newblock \bibinfo{title}{{Sensitivity of the high altitude water Cherenkov
  detector to sources of multi-TeV gamma rays}}.
\newblock \bibinfo{journal}{Astroparticle Physics}
  \bibinfo{year}{2013};\bibinfo{volume}{50}:\bibinfo{pages}{26--32}.
\newblock \DOIprefix\doi{10.1016/j.astropartphys.2013.08.002}.
  \href{http://arxiv.org/abs/1306.5800}{\tt arXiv:1306.5800}.
%Type = Inproceedings
\bibitem[{{Riviere} and {for the HAWC Collaboration}(2016)}]{Gamma_Colas}
\bibinfo{author}{{Riviere}\xfnm[ C.]}, \bibinfo{author}{{for the HAWC
  Collaboration}\xfnm[]}.
\newblock \bibinfo{title}{Preliminary hawc 1st year catalog}.
\newblock In: \bibinfo{booktitle}{Proceedings of the Gamma Symposium}.
  \bibinfo{year}{2016},.
%Type = Article
\bibitem[{{Sandoval} et~al.(2016){Sandoval}, {Lauer} and {Wood}}]{Atel1}
\bibinfo{author}{{Sandoval}\xfnm[ A.]}, \bibinfo{author}{{Lauer}\xfnm[ R.]},
  \bibinfo{author}{{Wood}\xfnm[ J.]}.
\newblock \bibinfo{title}{{HAWC detection of increased TeV flux state for
  Markarian 501}}.
\newblock \bibinfo{journal}{The Astronomer's Telegram}
  \bibinfo{year}{2016};\bibinfo{volume}{8922}.
%Type = Article
\bibitem[{{Biland} et~al.(2016){Biland}, {Dorner}, {Lauer}, {Wood}, {Kapanadze}
  and {Kreikenbohm}}]{Atel2}
\bibinfo{author}{{Biland}\xfnm[ A.]}, \bibinfo{author}{{Dorner}\xfnm[ D.]},
  \bibinfo{author}{{Lauer}\xfnm[ R.]}, \bibinfo{author}{{Wood}\xfnm[ J.]},
  \bibinfo{author}{{Kapanadze}\xfnm[ B.]}, \bibinfo{author}{{Kreikenbohm}\xfnm[
  A.]}.
\newblock \bibinfo{title}{{Enhanced and increasing activity in gamma rays and
  X-rays from the HBL Mrk421}}.
\newblock \bibinfo{journal}{The Astronomer's Telegram}
  \bibinfo{year}{2016};\bibinfo{volume}{9137}.
%Type = Inproceedings
\bibitem[{{Lenarz} and {for the HAWC Collaboration}(2016)}]{Gamma_GRB}
\bibinfo{author}{{Lenarz}\xfnm[ D.]}, \bibinfo{author}{{for the HAWC
  Collaboration}\xfnm[]}.
\newblock \bibinfo{title}{The hawc grb program}.
\newblock In: \bibinfo{booktitle}{Proceedings of the Gamma Symposium}.
  \bibinfo{year}{2016},.
%Type = Inproceedings
\bibitem[{{L\'opez-Coto} and {for the HAWC Collaboration}(2016)}]{Gamma_Ruben}
\bibinfo{author}{{L\'opez-Coto}\xfnm[ R.]}, \bibinfo{author}{{for the HAWC
  Collaboration}\xfnm[]}.
\newblock \bibinfo{title}{Hawc detection of very extended nearby pwne powered
  by old pulsars and their relation to the positrons at the earth}.
\newblock In: \bibinfo{booktitle}{Proceedings of the Gamma Symposium}.
  \bibinfo{year}{2016},.
%Type = Article
\bibitem[{{Sandoval}(2015)}]{HAWC_Outriggers}
\bibinfo{author}{{Sandoval}\xfnm[ A.]}.
\newblock \bibinfo{title}{{HAWC Upgrade with a Sparse Outrigger Array}}.
\newblock \bibinfo{journal}{ArXiv e-prints}
  \bibinfo{year}{2015};\href{http://arxiv.org/abs/1509.04269}{\tt
  arXiv:1509.04269}.
%Type = Article
\bibitem[{{P{\"u}hlhofer} et~al.(2013){P{\"u}hlhofer}, {Bauer}, {Eisenkolb},
  {Florin}, {F{\"o}hr}, {Gadola} et~al.}]{Flashcam}
\bibinfo{author}{{P{\"u}hlhofer}\xfnm[ G.]}, \bibinfo{author}{{Bauer}\xfnm[
  C.]}, \bibinfo{author}{{Eisenkolb}\xfnm[ F.]},
  \bibinfo{author}{{Florin}\xfnm[ D.]}, \bibinfo{author}{{F{\"o}hr}\xfnm[ C.]},
  \bibinfo{author}{{Gadola}\xfnm[ A.]}, et~al.
\newblock \bibinfo{title}{{FlashCam: A fully digital camera for the Cherenkov
  Telescope Array}}.
\newblock \bibinfo{journal}{ArXiv e-prints}
  \bibinfo{year}{2013};\href{http://arxiv.org/abs/1307.3677}{\tt
  arXiv:1307.3677}.

\end{thebibliography}

\end{document}